# Response to Scheuer-Yariv: "A Classical Key-Distribution System based on Johnson (like) noise – How Secure?" [1]

L. B. Kish

*Department of Electrical and Computer Engineering, Texas A&M University, College Station, TX 77843-3128, USA*

**Abstract.** We point out that the claims in the comment-paper of Scheuer and Yariv are either irrelevant or incorrect. The *idealized* Kirchoff-loop-Johnson-like-noise (KLJN) scheme is totally secure therefore it is more secure than idealized quantum communication schemes which can never be totally secure because of the inherent noise processes in those communication schemes and the statistical nature of eavesdropper detection based on error statistics. On the other hand, with sufficient resources, a *practical/non-ideal* realization of the KLJN cipher can arbitrarily approach the idealized limit and outperform even the idealized quantum communicator schemes because the non-ideality-effects are determined and controlled by the design. The cable resistance issue analyzed by Scheuer and Yariv is a good example for that because the eavesdropper has insufficient time window to build a sufficient statistics and the actual information leak can be designed. We show that Scheuer's and Yariv's numerical result of 1% voltage drop supports higher security than that of quantum communicators. Moreover, choosing thicker or shorter wires can arbitrarily reduce this voltage drop further; the same conclusion holds even according to the equations of Scheuer and Yariv.

The comment-paper by Scheuer and Yariv (Sch-Y) [1,2] attempts to show that the Kirchhoff-loop-Johnson-noise (*KLJN*) [2-8] classical communicator has limited security. Clarification of the issues of real security is indeed important especially because the KLJN system has recently became *network-ready* [6]. This new property [6] opens a large scale of practical applications because the KLJN cipher can be installed as a computer card [6], similarly to Eternet cards.

In this response, we focus on the most significant issues but additional arguments have been published in [9]. Sch-Y [1,2] claim that the first *KLJN* paper [3] has basic flaws and that the *KLJN* cipher is not secure. However, their arguments are incorrect and/or irrelevant [9], or already published [8]. Because these kind of mistaken claims about the security of physical secure layers arise due to mixing requirements of idealized cipher schemes with practical design issues, first we clarify what do idealized and practical security mean. We have recently published similar considerations in the paper about the natural immunity of the *KLJN* cipher against the *man-in-the-middle* attack [5]. Work is in hand about the mathematical analysis of the practical security and design aspects of the *KLJN* cipher [7]. In our response below we shall use some preliminary results of [7].

First of all, let us consider what *idealized (theoretical) total security* and *practical total security* of physical secure layers mean.

*i) Total security of the idealized model* [4]. This holds when the mathematical model of the idealized physical system shows an unconditional security. It means that the eavesdropper can extract zero information within the mathematical framework of the model; or if she is able to extract information, she disturbs the channel and she will be discovered. The KLJN cipher satisfies these conditions: in the exact mathematical model, the passive eavesdropper extracts zero information and the invasive eavesdropper extracts one bit of information until the eavesdropping is discovered [3,4].

---
[1] A more extensive version is submitted to arxiv.org on, February 1, 2006 and updated subsequently; http://arxiv.org/abs/physics/0602013

*ii) Total security of the practical situation* [4]. Since a real physical system is always more complex than idealized mathematical models, no practical system can be totally secure. For example, regarding quantum communication, ideal single-photon-source, noise-free channel and noise-free detectors do not exist and any of these non-idealities compromise total security. Still, we can talk about *total practical security* if the security can arbitrarily be increased by unlimited investment in the enhancement of the system. For example, the wire resistance in the KLJN system is also a non-ideality factor [8] however it can be arbitrarily reduced by using ticker wires or shorter connections, see below.

*iii) Information leak at idealized and practical situations*. In idealized quantum communication systems, information of the order of a few % of the number of transmitted bits can be extracted, for example if the eavesdropper randomly extracts a small fraction of photons [11,12], clone them (with ~70% fidelity), sends one photon back to the channel, and extracts the information from the remaining photon. This process will cause only a negligible change of the error rate in the channel so the quantum eavesdropper detection methods will not be able to detect the eavesdropping [10-12]. For that reason practical quantum communicators [12] must use *privacy amplification* technique [10-12], which is a software-based tool, extracting a short key with *guaranteed negligible information leak* from a long key with *possibly strong information leak*. Because the privacy amplification can be used in any secure communication system with any raw key, the fair comparison of the information leak of physical-secure-layer-type communicators requires the comparison of the information leaks of the raw bits. We will show that the KLJN cipher's information leak can easily be much less than that of idealized quantum communicators and this claim is supported even by Sch-Y's results when we analyze their implications. Now, let us deal with the main claims of Sch-Y in their comment paper.

**1.** In the first section Sch-Y state that our analysis in [3] *"contains a basic flaw"* because *"it completely ignores the finite propagation time"* between the sender and receiver. This statement is incorrect. The analysis in [3] is carried out in the so-called *quasi-static limit* of Maxwellian electrodynamics [13] guaranteeing that the *voltage and current along a wire are constant* and that excludes any propagation delay effects. This is expressed by Eq. (9) in [3] and the related text:

$$f_{max} L \ll c \qquad (1)$$

where $f_{max}$ is the highest frequency component of the voltage and current, $L$ is the cable length (range of communication) and $c$ is the propagation velocity. *This equation implies not only a necessary condition for the security but also the necessary condition that our Kirchoff-loop circuit model, the idealized mathematical model of the cipher, holds* because this condition is the base of the theory of electronic circuits with discrete components. With this criticism, Sch-Y have missed to realize that the high-frequency limit posed by Eq. (1) in the *frequency domain* is equivalent to the quasi-static limit in the *time domain*. Transformation of limits between the time and frequency domain are basic and well-known tools due to Fourier-theory so the reader was supposed to have and imply this knowledge.

**2.** In Section 2, Sch-Y analyze the limit of high-frequency bandwidth (short correlation time) of noise. This limit was excluded in [3] by the Eq. (1) shown above because then the KLJN cipher naturally fails to function; therefore Sch-Y's analysis is irrelevant for [3]. Fortunately, Sch-Y also arrive at the same conclusion that the KLJN cipher is useless in this limit, thus there is no contradiction between Sch-Y [1,2] and [3] in this respect.

**3.** At the beginning of Section 3 of [1,2], Sch-Y claim rightly that if the ciphers use fast switches then propagation time effects will still occur, even with low-frequency noise, and the cipher will be non-secure. Though it is correct, this claim is *again irrelevant* because due to Fourier-theory, it violates directly Eq. (1) shown above, so this situation was excluded in [3]. As it is well-known and mentioned above, short time scale and fast time processes in the time domain correspond to high frequency scale and high frequency components in the frequency domain consequently fast switching and their transients produce high-frequency components in the channel. Therefore, a practical realization of the KLJN cipher will naturally need line filters, which are low-pass filters at the two ends of the line [4], see Figure 1. This is also necessary to defend against possible attacks by high-frequency probing signals. Of course, these are all practical realization problems [4,7] and as such they were out of the scope of [3].

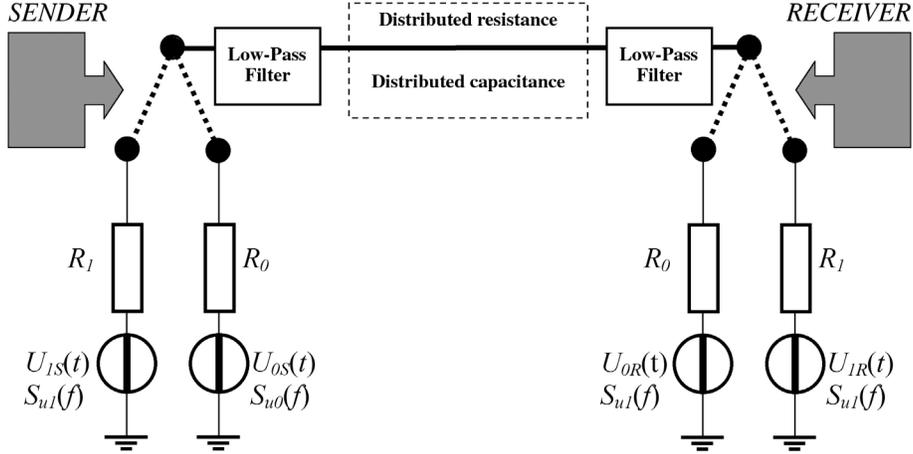

**Figure 1.** Toward the modeling and realization of practical KLJN ciphers.

**4.** Finally, in the rest of Section 3 of [1,2], Sch-Y analyze the practical problem of non-zero wire resistance. Note, this security problem was first pointed out by Janos Bergou [8]. At certain wire length and specific resistance, their numerical results for the relative difference of the mean-square (MS) voltages at the two ends of the line, in the case of secure bit exchange, is 1%. Then Sch-Y claim that this 1% drop of the MS voltage *"can easily be detected and allow Eve to determine Alice and Bob's selection of resistors"*. Though we accept the 1% drop of the MS voltage as a realistic practical goal [7] we disagree with Sch-Y's claim that the eavesdropper can easily detect this 1% drop.

With the very same voltage drop, we have carried out a model study [7] of the distribution functions of the voltages, currents and the drop of the MS voltage for $R_1 / R_0 = 10$, with a *linear full-wave detector* [14] and clock period

$$\tau_c = 3 / f_{max} . \qquad (2)$$

Eq. (2) results in *relative standard deviation* 0.2 of the voltage and current statistics [14]. The results are summarized in Figure 2. Note, only the relative positions and the shape of the curves have meaning, not the actual x and y values. During the clock period, due to Eq. (2), the time is enough only for a few statistically independent sampling of these distribution functions. This sampling is enough for the sender and the receiver, see Figure 2 (a), to decide between the two functions with 0.3% error rate [7]. However the eavesdropper, who measures the voltage drop, has to decide between the two situations by sampling the f(x) and g(x) density functions given in Fig. 2 (b) and that must be done with the same small number of independent samples. The *characteristics width* (standard deviation) of these curves (20% of the peak's x coordinate) is 20 times greater than the difference of the locations of the x coordinates of the peaks (1%). The eavesdropper's task seems to be hopeless by the naked eye however, by using proper statistical tools, she can still extract some information. A deeper analysis based on *Shannon's channel coding theorem* [7] concludes that in this case the *upper limit of information leak* is 0.7% of the transmitted bits. This is close to but less than the information leak of quantum communicators without privacy amplifier software (see above). Thus Sch-Y's 1% drop of the MS voltage yields a lower information leak than that of quantum communicators.

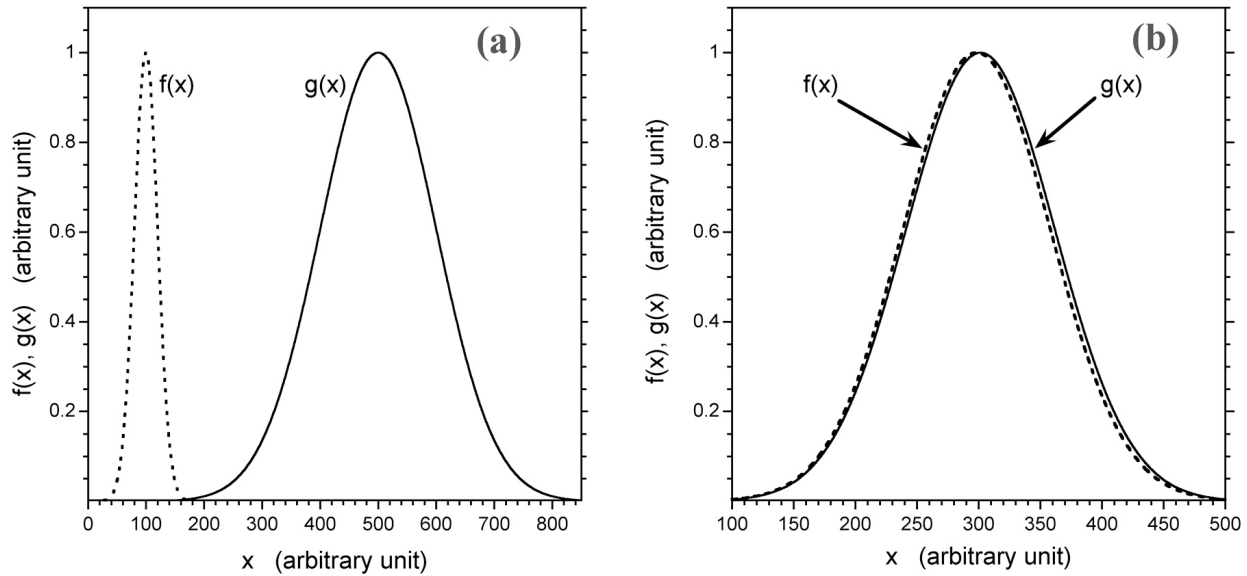

**Figure 2.** Model study of distribution functions [7]. (a): Amplitude distribution functions sampled by the sender and receiver. (b): Amplitude distribution functions sampled by the eavesdropper at the two ends of the wire.

To illustrate how the security can arbitrarily be increased, let us suppose that we increase the wire diameter by a factor of 10. Then, according to Sch-Y's Eq. (12), the relative voltage drop will decrease by a factor of 100 which makes the difference of the location of the x coordinates of the peaks 100 times less than it is presently in Figure 2 (b). This is an impressive improvement making it virtually impossible for the eavesdropper to extract useful information. These considerations illustrate why the KLJN cipher can be designed so that its practical realization is much closer to the limit of *unconditional security* than even the idealized quantum communicators.